# CELL DISSOCIATION :
# A COHESIVE FORCE OF HYDRODYNAMIC ORIGIN

H. Vasseur.

PSC, 33 rue St Leu, 80039 Amiens Cedex, France.
(contact : hugues.vasseur@u-picardie.fr)

*When an experimentalist or a biological mechanism applies an external force onto a cell chemically sticking to its substrate, a reacting "suction" force, due to the slow penetration of the surrounding fluid between the cell and the substrate, opposes to the dissociation. This force can overcome other known adhesive forces when the process is sufficiently violent (typically $10^5 pN$). Its maximal contribution to the total adhesive energy of the cell can then be estimated to $2 \; 10^{-3} J/m^2$. The physical origin of this effect is quite simple, and it may be compared with that leaning a "suction-cup" against a bathroom wall. We address the consequences of this effect on (i) the dissociation energy, (ii) the motion of the fluid surrounding the cell, more especially, on the pumping of the fluid by moving cells, and (iii) the inhibition of cell motion.*

**Introduction**: The phenomenon of cell adhesion is fundamental in cell functioning and in various inter-cell metabolic activities of biological tissues [7]. For instance, cell division [1], cell differentiation [2], cell migration (lamellipode) [3], infections (adhesion of pathogenic agent) [4], leucocytes/endothelium interaction [4], and colonization by the cells of a primitive cancerous tumor [5] are controlled by sticky interactions between cells and their environment. These interactions are due to organic molecules, a hundred of which having yet been characterized [6]. An important stage for understanding these phenomena has been got over by Bell [8] when he described the dissociation kinetics of proteins diffusion. On pointing out the formation of links between a cell and a substrate, his results have stimulated a number of works on link properties, dynamics of the cells/sticky-molecules dissociation, characteristic times [9-11], sticky force strengths [12-13] and the so called "breaking force" [14]. They have shown in particular that the survival time for links between Streptavidin-Biotin (archetype of bond between ligand and receptor) varies between $10^{-3}$sec and 60sec when the loading rate of the external force varies from 60000pN/sec to 0.1pN/sec, whereas the sticky force strength for a bond typically varies between 1pN and 1000pN [11]. Such a wide range of time and force scales opens the possibility of coupling the cell motion to various process occurring in biological tissues, in particular hydrodynamic phenomena.

Along this way, we show in this letter that an *additional* force of hydrodynamic origin plays an important role in the cell/substrate dissociation dynamics, which has never been analyzed within the context of cellular adhesion [15]: An experimentalist or a biological mechanism applying abruptly an external force on the cell for separating it from its substrate induces a bonding force which opposes to the separation. The inhibition of the surrounding liquid motion is at the origin of this force. Its effect is very similar to that leaning a suction-cup against a bathroom wall and will be, accordingly, denoted by "suction force" below. Indeed, when the cell begins to be pulled out, the pressure under the cell diminishes, which leads to penetration of the surrounding liquid between the cell and its substrate. The corresponding pressure difference between the top and the underside of the cell generates the suction force. Thus, in contrast to similar hydrodynamic forces caused, for instance, by shear flow [16, 17], suction is purely attractive.

So, the cell separation is prevented by the links, on the one hand, and by suction, on the other hand. When the external force is large enough the links are finally broken, and we will show that this occurs simultaneously with the disappearance of suction. For a cell whose initial contact radius is $5\mu$m the maximum suction force is $\approx 17\ 10^4$pN. It cancels progressively during the fluid penetration and becomes negligible when all the links are broken. In fact, when the raising velocity of the cell is low, the actual suction force is much smaller than $17\ 10^4$pN. This maximal value is reached only when the velocity overcomes a critical velocity, denoted by $V_c$ hereafter. This critical velocity permits us to distinguish two dynamical regimes called under and over critical, respectively.

**Suction effect**.
- Under-critical regime: $V<V_c$.

Let us first study the case when the raising velocity V of the cell underside surface is small. When the cell underside is raising, $P_2$ under the cell is lower than the pressure $P_1$ of the surrounding fluid. The difference of pressure, $\Delta P=P_1-P_2$, corresponds to an horizontal gradient (parallel to the substrate), on the one hand, and a "vertical" jump (between the bottom and the top of the cell), on the other hand (see Fig. 1-a). Thus, it has two connected effects: (i) It generates an horizontal fluid flow towards the center of the cell underside. The liquid follows then the ascending motion of the cell, yielding a pumping of the external fluid towards the cell-substrate contact zone. (ii) A vertical suction force tackling the cell against the substrate, caused by the difference of pressure between the top of the cell ($P_1$) and its underside ($P_2$).

Thus, $\Delta P$ is related to both the liquid flow and the suction force. At sufficiently small velocity (under-critical regime) $P_2$ is strictly positive and decreases with V. At the minimum value $V_c$ of V such that $P_2=0$, the regime becomes "critical". In this case, the pressure of the fluid just below the cell vanishes, but no vacuum zone can be created between the fluid surface and the cell underside.

- Over-critical regime: $V>V_c$.

$V_c$ may also be regarded as the maximum speed of the underside surface for which this surface remains in contact with the liquid. At larger velocities (over-critical regime) the fluid is no longer able to fill the volume liberated by the cell displacement. Indeed, if the raising velocity is larger than $V_c$, then the fluid fails to remain in contact with the cell surface, because the vertical speed of the fluid surface under the cell remains locked to its limit value $V_c$. A vacuum zone with $P_2=0$ (in fact, low pressure gas) would then be provisionally created between the cell and the fluid, the volume of the entering fluid being smaller than the volume freed by the raising underside surface. We shall not study further the situation where V is larger than $V_c$ since the existence of a vacuum zone has yet never been reported. However, it is worthwhile to study the critical regime $V=V_c$ since the suction force and energy are then maximum.

**Separation time $\Delta \tau^c$ in the critical regime**:

It may be seen that, since the cell membrane is deformable, in a realistic model of separation the suction force should vary with time and position below the cell. It is evident that suction is negligible when the fluid can easily penetrate under the cell. Thus, the suction force is mainly concentrated where the entrance of the fluid is strongly inhibited by : (i) The presence of unbroken links which constitute a barrier against fluid penetration, and (ii) the narrowness of the layer between the cell and the substrate. This defines a small area located around a closed line moving from the border (at the beginning of the separation process) towards the center of the cell. In the region between the center and the line, the cell is still tackled and the links are at rest, whereas between the line and the cell border the links are already broken. When this "separation line" (Fig. 1-c,d), where the raising process just begins, has reached the center, all the links are broken and the cell is free.

Similarly, the velocity V is also not constant. The exact form of the interface and the time and space variations of V depend of course on the way the external force is applied and on the elastic properties of the cell, which are yet far from being known (see, for instance, Ref. [13] in the discussion about eukaryotic cells). The suction force depends crucially on these unknown properties and could be calculated in detail only within a global theoretical model of cell deformation. However, the fact that the suction effect takes place mainly in the small area around the separation line permits us to evaluate most of its properties without referring to the applied force and using a simplified cell deformation model.

So, in order to simplify the estimation of $\Delta \tau^c$ and $V_c$ (which, from its definition, depends also on position and time!), we assume that the sticky molecules are approximately distributed on a square network with lattice spacing $\varepsilon$. Moreover, we consider the initial contact surface of the cell with the substrate as a square with side length $n\varepsilon$ which is constituted of independent rigid parts having the form of concentric square coronas represented in

Fig. 1-c. The corona i (the outside corona is labeled by i=1) has a perimeter $L_i=4(n-2i+2)\varepsilon$, and an area $S_i=4(1-2i+n)\varepsilon^2$. At t<0 all the coronas are tackled against the substrate. When the external force is applied at t=0, the outside corona, i=1, begins to rise and the fluid begins to penetrate under the cell. We denote by $h_i$ the height of the cell above the substrate at the external edge of the corona i (it depends only on i because we assume a symmetric flow).

Let us first describe the way the liquid enters under the cell. Two neighboring sticky links form, together with the cell surfaces and the substrate, a "door" by which the fluid enters into a cavity with volume $h_i\varepsilon^2/2$. Such door and cavity are represented in Figs. 1-b,c,d. The pressure is almost constant inside the cavity and varies only across the door. The difference of pressure $\Delta P$ between both sides of the door provokes the fluid motion towards the cavity. The door plays the role of a classical hydrodynamic pipe parallel to the substrate (its section being, for simplicity, considered as almost elliptic) crossed by a flow of fluid q given by the classical formula [19]:

$$q = \frac{\pi}{64\eta l} \frac{h_i^3 \varepsilon^3}{h_i^2 + \varepsilon^2} \Delta P \qquad (1)$$

where the length l of one pipe is equal to the diameter of the link section (Fig. 1-b), $\eta$ is the dynamic viscosity of the fluid, $h_i/2$ (half height of the sticky link) and $\varepsilon/2$ being the principal axes of the ellipse. In this equation $\Delta P$ can be considered either as the cause of the fluid motion or as its consequence giving rise to the suction effect. Let us notice that, before stretching the sticky links by the applied force, the height $h_f$, which separates at rest the cell from its substrate, is not strictly zero because of the presence of finite sized sticky molecules and thermal fluctuations of the cell membrane. Typically, $h_f$ can be estimated to 4 nm [21]. Thus, the fluid is already present under the cell before the stretching process is activated, so that, in contrast with an actual suction cup, there is no force when the cell is at rest.

While the cell is progressively raised, the fluid penetrates from outside towards the center of the contact surface with the substrate. At the level of the corona i, its flow $Q_i$ is proportional to the number of doors it contains:

$$Q_i = q \frac{L_i}{\varepsilon} \qquad (2)$$

When $h_i<<\varepsilon$ the role of the links in the inhibition of the fluid motion is negligible and the flow $Q_i$ in Eq. (2) depends no longer on $\varepsilon$ and becomes proportional to $h_i^3$. This arises, for each corona, at the beginning of the raising process, or when the density of sticky links is small. The fluid penetration is then slowed down only by the smallness of $h_i$, and it increases strongly with the height of the cell. When $h_i>>\varepsilon$, the barrier to the motion due to the doors width $\varepsilon$ becomes efficient and the flow $Q_i$ increases more slowly with time and varies only as $h_i\varepsilon^2$. At this step of the process large densities of sticky links much inhibit the fluid flow and increases the efficiency of the suction effect.

Let us now turn to the cell response. We assume a simple zipper-like model for describing its deformation [25]: Once the $(j-1)^{th}$ corona height $h_{j-1}$ reaches a critical value $h_c$, the neighboring corona j begins to rise in its turn. Figure 1-d shows the profile of the cell during the raising process. This is equivalent to assume that the membrane cannot be bent by an angle larger than a critical value characteristic of its local elasticity. The critical height is to be compared with the "breaking" length $h_B$ above which a link is broken [20]. It turns out that $h_B$ and $h_c$ take usually close values. We will see in the discussion of this letter that when $h_i>h_c$, the flow is almost instantaneous so that there is always only a single corona around the separation line in which suction is effective: When a new corona begins to raise, the previous one is liberated. We denote the former by "border corona".

The critical velocity $V_{ci}$ (see Fig. 3) at the level of the border corona i is, by definition, the cell membrane velocity above the corona when $\Delta P=P_1$:

$$V_{ci} = 2\frac{Q_i(t)}{S_i} \quad (3)$$

Integrating the balance equation $Q_i=qL_i/\varepsilon=d\Omega_i/dt$ (where $\Omega_i=1/2(h_i-h_f)S_i$ is the volume of the fluid penetrating above the corona i) yields:

$$(\frac{4\eta l}{\varepsilon^2 \pi \Delta P L_i})8\ [S_i Log(h_i S_i) - \varepsilon^2 S_i^5 \frac{h_i^2}{2}] + A = t \quad (4)$$

where A is a constant of integration. Our calculation shows that when $h_i$ becomes very large the fluid penetration becomes almost instantaneous (see Fig. 2). Accordingly, as stated above, the suction effect disappears in a given corona as soon as it is liberated from the substrate. The separation time above the corona i reads then:

$$\Delta \tau_i^c = t(h_i=h_c) - t(h_i=h_f) \quad (5)$$

where $t(h_i)$ is given by the left hand side of Eq. (4). The separation time of the whole cell is therefore given by: $\Delta \tau^c = \sum_{i=1}^{N} \Delta \tau_i^c$, where $N=(n+1)/2$ when n is odd, and $N=n/2$ for n even.

In the critical regime the suction force depends simply on time. Indeed the pressure exerted on the border corona is constant, so that the force depends only on its decreasing area. The suction force amounts thus $4(1-2i+n)\varepsilon^2 P_1$. It decreases when i increases, i.e., when t increases, between a maximum value, $4(n-1)\varepsilon^2 P_1$, and zero during $\Delta \tau^c$ (Fig. 4).

**Separation time $\Delta \tau$ in the under-critical regime**: In under-critical conditions one has $V_i<V_{ci}$, so that $\Delta \tau$, which is then larger than $\Delta \tau^c$, is given by: $\Delta \tau = \sum_{i=1}^{N} \Delta \tau_i$, where $\Delta \tau_i$ is such that $h_c = \int_0^{\Delta \tau_i} V_i(t)dt$. The corresponding suction pressure varies now with time since $P_2(t)$ is no longer locked to zero by the criticality condition. For each corona $P_2(t)$ can be deduced from the equations $Q_i(t)=V_i(t)S_i/2=q(t)L_i/\varepsilon$, and $h_i(t)=h_f+\int_0^t V_i(t')dt'$:

$$\Delta P(t) = \frac{32\varepsilon \eta l}{\pi}\ \frac{S_i}{L_i}\ \frac{h_i(t)^2+\varepsilon^2}{\varepsilon^3 h_i(t)^3}\ V_i(t) \quad (6)$$

The work of the suction force ("suction energy") before the separation is achieved is given by [20]:

$$W_S = \sum_{i=1}^{N} S_i \int_{h_f}^{h_B} \Delta P(h_i(t))dh_i \quad (7)$$

We have seen that dissociation is a quite complicated process, which depends on the unknown elastic properties of the cell, elastic and plastic behaviors of the links, and on the way the external force is applied onto the system. Therefore, we are not able to reliably predict in detail the behavior of $V_i(t)$ in any realistic situation. However, in order to get an order of magnitude of the suction energy, let us consider $V_i(t)$ as practically independent of time for the border corona. Equation (6) shows then that the suction force decreases with time. Inserting Eq. (6) into Eq. (7) gives:

$$W_S = \frac{64\eta l \varepsilon}{\pi}V\{2Log(\frac{h_B}{h_f})+\varepsilon^2\frac{h_B^2-h_f^2}{h_f^2 h_B^2}\}\ \kappa(n) \quad (8)$$

where $\kappa(n) = \sum_{i=1}^{N}(n+1-2i)^2/(n+2-2i)$ is a numerical factor varying between 1/2 and $N^2$ as N becomes very large.

**Discussion**: The suction force and its contribution to the cell/substrate separation energy are not negligible with respect to those reported in the literature [11,13] for chemical links. On using Eq. (3) and Eq. (5) we can estimate numerically the raising velocity $V_i$ of the cell underside (see Fig. 3) and the lifetime $\Delta\tau^c$ of the separation process in the critical regime (numerical values are calculated from experimental data reported in Ref. [13] or, otherwise, the parameters will be specified): For $h_c=2.4\ 10^{-8}$m [18,20], we find $\Delta\tau^c=4.2\ 10^{-5}$sec. This value is almost independent of $h_c>2.4\ 10^{-8}$m ($\Delta\tau^c=4.5\ 10^{-5}$sec for $h_c=10^{-6}$m, $\Delta\tau^c=4.3\ 10^{-5}$sec for $h_c=5.4\ 10^{-8}$m, $\Delta\tau^c=4.2\ 10^{-5}$sec for $h_c=2.4\ 10^{-8}$m, $\Delta\tau^c=3.9\ 10^{-5}$sec for $h_c=1.4\ 10^{-8}$m, $\Delta\tau^c=3.4\ 10^{-5}$sec for $h_c=9\ 10^{-9}$m, $\Delta\tau^c=2.8\ 10^{-5}$sec for $h_c=7\ 10^{-9}$m). Indeed, above $h_i>2,4\ 10^{-8}$m, the fluid penetration becomes so fast (from 0.5m/sec. to several m/sec) that it can be considered as almost instantaneous whatever the value of $h_c$. This justifies that we have considered the suction as negligible when $h_i(t)>h_c$ in the previous calculation. On the other hand, Eqs. (4,5) show that $\Delta\tau^c$ depends on $h_f$, the viscosity $\eta$, the cell size and the links thickness $l$. In particular, it considerably decreases when $h_f$ increases. For instance, $\Delta\tau^c=6.7\ 10^{-4}$sec for $h_f=1$nm. The suction effect is reinforced as $h_f$ get smaller. More specifically, if one set $l=30$nm, $h_f=4$nm and $h_c=2.4\ 10^{-8}$m, then $\Delta\tau^c=42\ 10^{-5}$sec. This order of magnitude of $\Delta\tau^c$ does not lie within the time scale (from $10^{-3}$ to several seconds) studied in Ref. [13] and, more generally, in the specialized literature. So, it is not surprising that the suction effect has not been identified yet.

The maximal separation energy for a total initial contact area $S=8.1\ 10^{-11}$m$^2$ is $\approx 2\ 10^{-3}$J/m$^2$, and the maximal suction force is $1.7\ 10^5$pN (see Fig. 5-c). In the under-critical regime, these values decrease. For instance, when $V=10^{-3}$m/sec for each corona, $W_S/S=6\ 10^{-5}$J/m$^2$, i.e., $W_S=3.8\ 10^{-15}$J for $S=8.1\ 10^{-11}$m$^2$. The corresponding suction force and pressure strength are plotted for each border corona when $h_i=2.4\ 10^{-8}$m in Fig. 5-a, and when $h_i=h_f$ in Fig. 5-b (between $h_i=h_f$ and $h_i=h_c$ $\Delta P_i(t)$ decreases and $\Delta P(t)$ oscillates). The total separation energy contains in addition a contribution necessary to stretch and break the sticky links, the maximum of which being $8\ 10^{-5}$J/m$^2$ [13]. Unfortunately, the raising speeds are not given in Ref. [13], so that the comparison with the predictions for the suction effect is difficult. One sees that the link contribution is one order of magnitude smaller than the suction energy in the critical regime, and one order of magnitude larger than in the under-critical regime for $V=10^{-3}$m/sec.

Suction takes place even in systems in which the cell separation is not necessarily described by the zipper model. Consider, for instance, two cardiac cells glued together by sticky links (desmosomes) with contact area $10^{-12}$m$^2$ and $h_B=10$nm. The presence of desmosomes (sticky links assembled in rigid plates) between the cardiac cells prevents zipper-like separation, and the cells are stretched without deforming the contact zone. Hence, the order of magnitude of the suction force and energy in the critical regime can be estimated to $10^5$ pN and $10^{-3}$J/m$^2$, respectively. Along the same way, conjonctive tissues are made of different fibers embedded in water, which confer their elastic properties. The water flow participates to these properties by a mechanism analog to the suction, except that the deformation happens without tearing. If one rapidly stretches a part of the tissue on a area equal to $10^{-6}$m$^2$, the maximal suction force and energy in the critical regime are then equal to $10^{-1}$N and $10^{-7}$J for $10^{-6}$m stretching.

The suction opposes to separation as well as, more generally, to any cellular fluctuation. In fact, it dissipates a part of the metabolic energy produced by the cells for generating small movements around their equilibrium positions in biological tissues. At this point of view, the suction plays an active role of regulator. This regulation can be estimated when one knows the amplitude and the frequency of the cell motion. Unfortunately, these data are usually not known for in-vivo cell vibrations. However, their order of magnitude can be deduced from data reported in Ref. [21], concerning cell wall oscillations in yeast cells with $5\mu m$ diameters surrounded by air. The amplitude of the wall vibrations is 3nm, with a mean velocity V=2.6 $10^{-6}$ to 4.9 $10^{-6}$m/sec. The maximum internal force and energy that the cell metabolism can generate are given by the authors of the reference: $10^{-8}$N and 3 $10^{-17}$J during one-oscillation with 3nm amplitude. Considering now the same cell linked [18] to a substrate permits us to estimate the energies dissipated by suction when the fluid is either air or water: (i) $W_S$=3.7 $10^{-20}$J and $W_S$=2 $10^{-18}$J, respectively, when V= 2.6 $10^{-6}$m/sec ; (ii) 7 $10^{-20}$J and 3.9 $10^{-18}$J, respectively, when V=4.9 $10^{-6}$m/sec. One sees that, at these velocities, the suction would use a negligible part of the metabolic energy. Nevertheless, the suction effect might act as a regulator of the cells fluctuations to prevent large amplitudes or velocities. Indeed, with the previous amplitude in water the velocity of the wall can not reach 3.8 $10^{-5}$m/s because the whole metabolic energy would be dissipated by suction. By the previous regulation effect, the suction participates to the restriction of the nutriments (or dangerous elements) pumped by the cell in its environment [21]. More generally, suction might regulate the intercellular fluid flow (in a similar way the blood flow is regulated by the metabolic activity for optimizing the oxygenation of cells [22, 24]). A failure of this regulation, provoked by a modification of the suction parameters could then participate to various diseases.

We have seen that in general $\Delta\tau^c$ is shorter than typical separation times (from $10^{-3}$ to several seconds) reported in the literature for artificial as well as natural inter-cell motions. For such velocities the dynamics is under critical, and the suction energy barrier is smaller than the sticky one. On the contrary, when considering violent processes, which can be obtained under extreme external conditions (e.g., shocks, tears, etc…), the suction effect becomes the dominant cohesive factor of the cell assembly. Unfortunately, such phenomena have not yet been studied experimentally at the relevant time scales. Sharpened studies of violent processes at very small time could reveal new and unexpected phenomena and could then give new insights into the organic system under extreme stress. Let us finally note that the suction could be considerably magnified if cavitation-type effects would take place in the under-cell liquid. Indeed, in this case metastable negative pressures $P_2$ are possible [26] and appears at values of $V/V_c$ larger than unity, which would lead to a significant increase of the maximal energy barrier. Although such cavitation effects have not yet been reported with usual low-viscosity organic fluids, one expects very large suction energies, even at low velocities (because, in addition, the viscosity diminishes $V_c$), at least when the biological fluids are more viscous than water.

**Acknowledgements :** I acknowledge stimulating discussions with B. Mettout, P. Nassoy, C. Gay, J.F. Joanny, Ri. Bouzerar, J.P. Morin, and A. Cherqui.

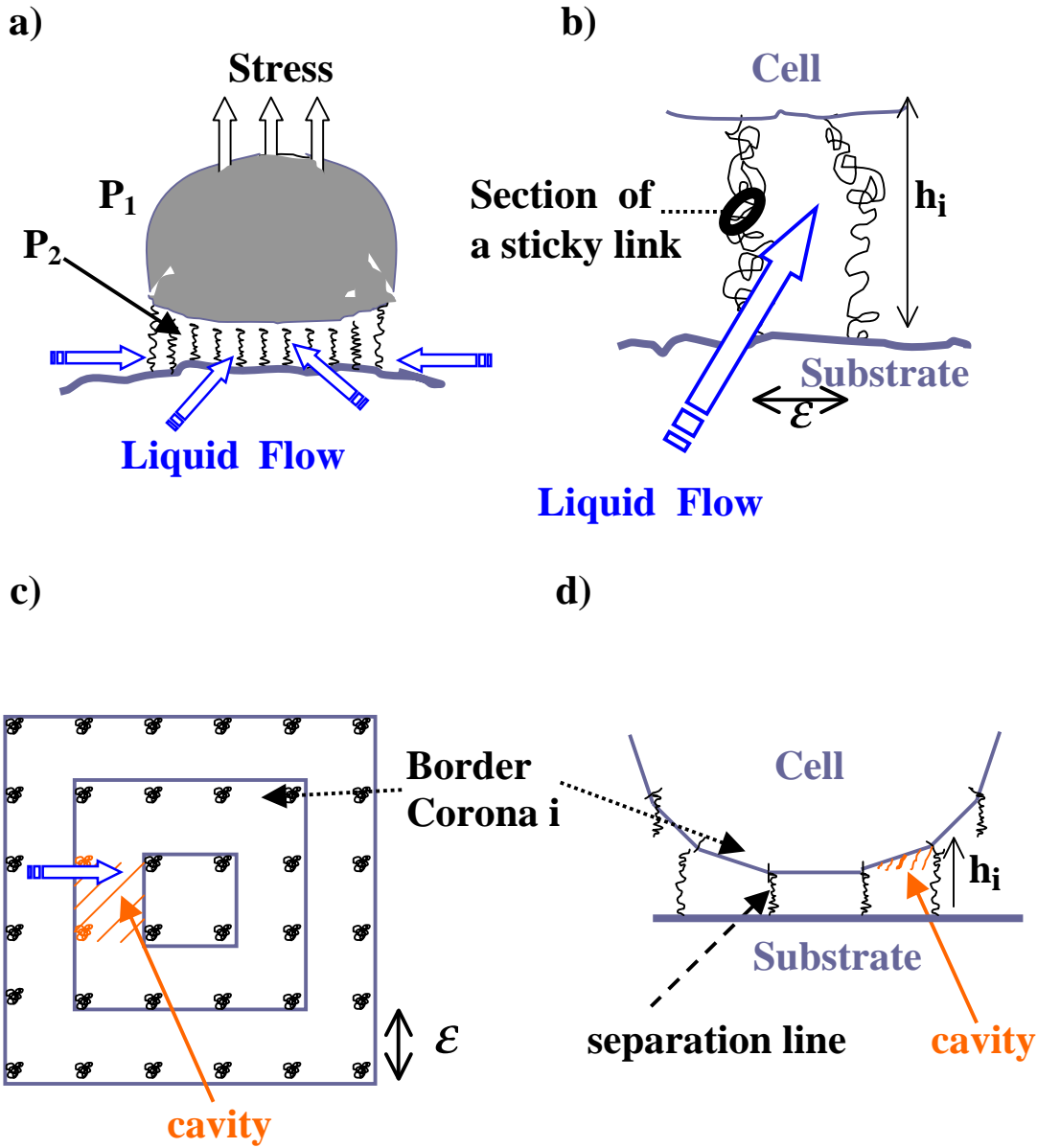

Fig. 1 : (a) Scheme of a cell in a liquid at pressure $P_1$. The links sticking the cell to its substrate are represented by small springs. When the cell is submitted to an external stress the links are stretched and the liquid penetrates below the cell. (b) Section of a door limited by two sticky links. (c) Square network formed by the links seen from above. It forms coronas which raise successively during the cell/substrate separation process. (d) Profile of the cell bottom. Each segment on the figure represents the section of a corona. At each time a single corona exhibits non-negligible suction, it surrounds the separating line. The liquid fills progressively the corresponding cavities while the volumes above external coronas fill up instantaneously. In (c) and (d) the links in the central "corona at rest" are not yet stretched. The links in the intermediate "border corona" are stretched but not broken, whereas the links of the external "liberated corona" are already broken.

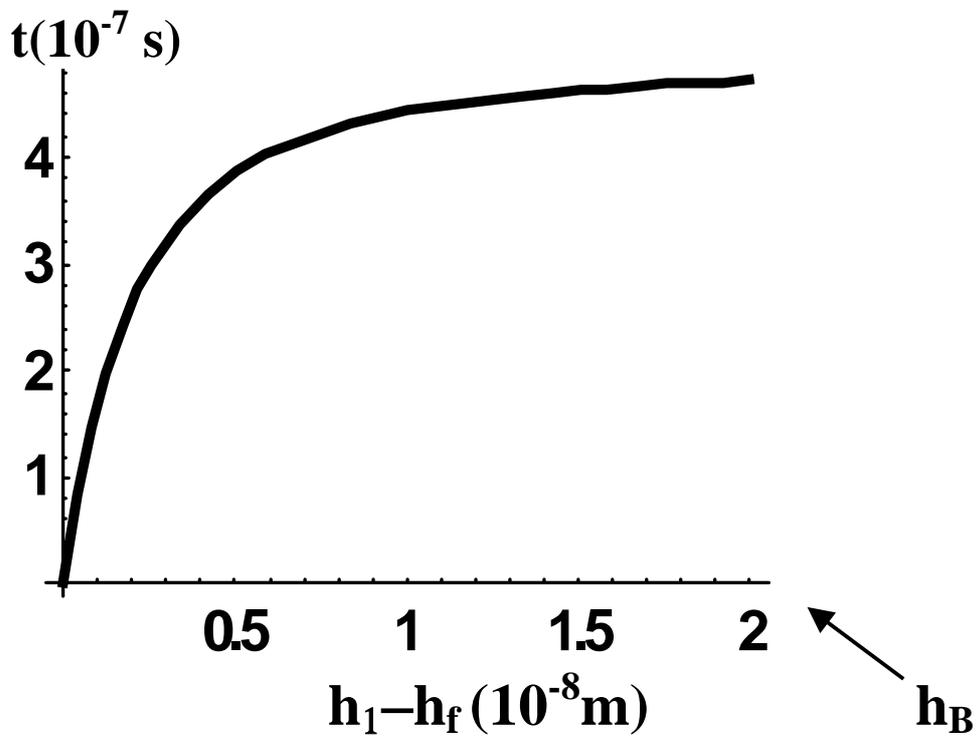

Fig. 2 : Time variation of the height $h_1$ of the cell membrane above the most external corona $i=1$ of the cell. $h_f=410^{-9}=$, $h_B=h_c=2.4 \cdot 10^{-8}$m.

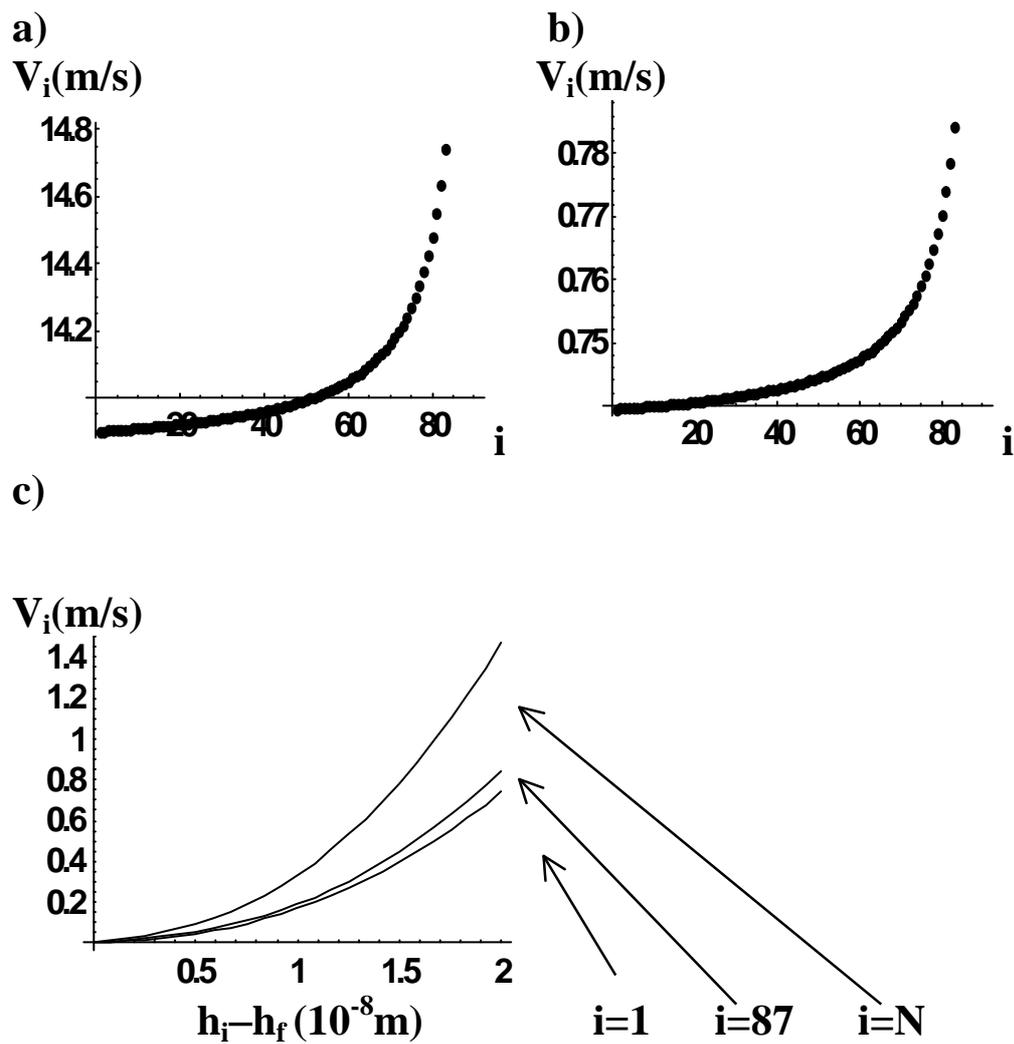

Fig. 3 : Raising velocity $V_i$ vs. the index corona i in the critical regime ($V_i=V_{ci}$). (a) when $h_c-h_f=10^{-7}$m. (b) when $h_c-h_f=2 \ 10^{-8}$m. (c) $V_i$ vs. $h_i$ for i=1, 87 and 90=N. The relation $V_{90}>V_{87}>V_1$ is due to the fact that the total number of doors in a corona varies with i.

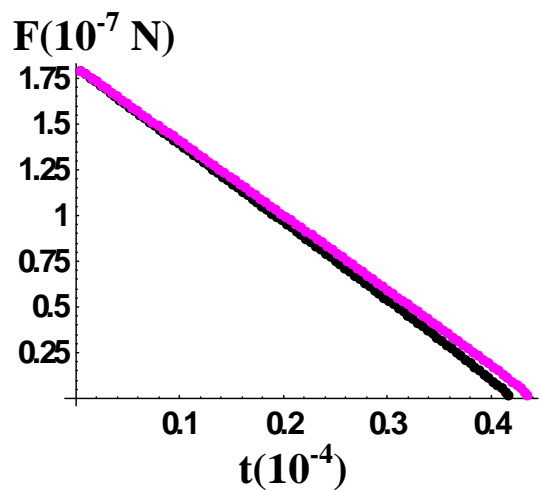

<u>Fig. 4</u> : Total suction force F vs. *t* in the critical regime for $h_c=10^{-7}$m (black dots) and $h_c=2,4 \cdot 10^{-8}$m (pink dots). The decreasing of F comes from the fact that the corona area diminishes when *i* increases.

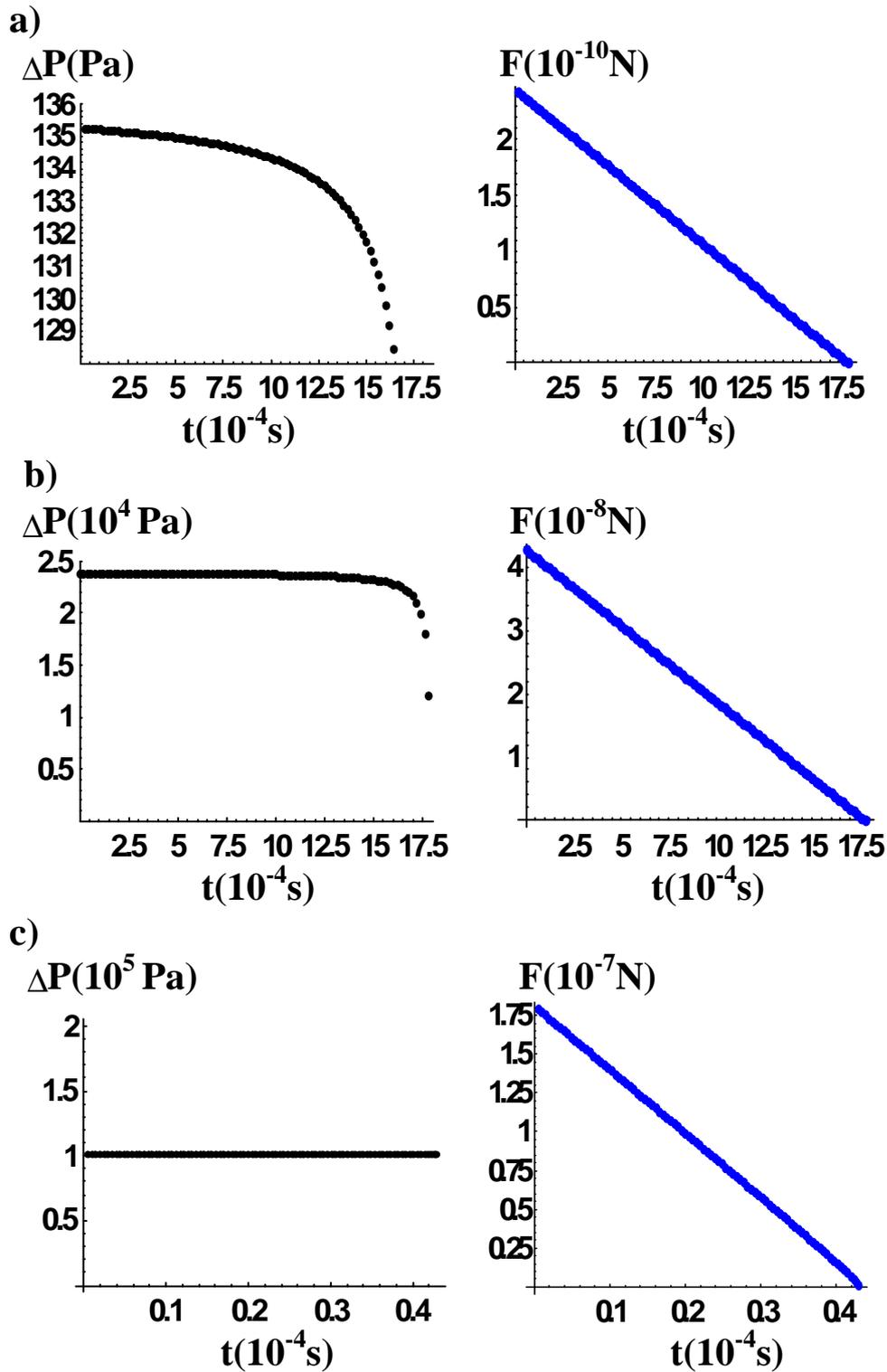

Fig. 5 : Suction pressure $\Delta P$ and suction force F in the under-critical (a and b) and critical (c) regimes (with $V_i=10^{-3}ms^{-1}$, $h_f=4\,10^{-9}m$, $h_B=h_c=2.4\,10^{-8}m$). (a) For each corona F and $\Delta P$ are calculated at the moment when $h_i=2\,10^{-8}m$. (b) At the moments when $h_i=h_f$. Although $V_i$ is the same for each corona, $\Delta P$ is not constant because the number of "doors" by unit of surface depends on i. (c)F and $\Delta P$ in the critical regime